\documentclass[aps,prl,twocolumn,superscriptaddress,showpacs,preprintnumbers,amsmath,amssymb]{revtex4}
\usepackage{epsfig}

\begin{document}

\title{\large
Measurement of $|V_{ub}|$ using inclusive $B \rightarrow X_{u} \ell \nu$ 
decays with a novel $X_u$-reconstruction method}

\affiliation{Budker Institute of Nuclear Physics, Novosibirsk}
\affiliation{Chiba University, Chiba}
\affiliation{University of Cincinnati, Cincinnati, Ohio 45221}
\affiliation{Gyeongsang National University, Chinju}
\affiliation{University of Hawaii, Honolulu, Hawaii 96822}
\affiliation{High Energy Accelerator Research Organization (KEK), Tsukuba}
\affiliation{Hiroshima Institute of Technology, Hiroshima}
\affiliation{Institute of High Energy Physics, Chinese Academy of Sciences, Beijing}
\affiliation{Institute of High Energy Physics, Vienna}
\affiliation{Institute for Theoretical and Experimental Physics, Moscow}
\affiliation{J. Stefan Institute, Ljubljana}
\affiliation{Kanagawa University, Yokohama}
\affiliation{Korea University, Seoul}
\affiliation{Kyungpook National University, Taegu}
\affiliation{Swiss Federal Institute of Technology of Lausanne, EPFL, Lausanne}
\affiliation{University of Ljubljana, Ljubljana}
\affiliation{University of Maribor, Maribor}
\affiliation{University of Melbourne, Victoria}
\affiliation{Nagoya University, Nagoya}
\affiliation{Nara Women's University, Nara}
\affiliation{National Lien-Ho Institute of Technology, Miao Li}
\affiliation{Department of Physics, National Taiwan University, Taipei}
\affiliation{H. Niewodniczanski Institute of Nuclear Physics, Krakow}
\affiliation{Nihon Dental College, Niigata}
\affiliation{Niigata University, Niigata}
\affiliation{Osaka City University, Osaka}
\affiliation{Osaka University, Osaka}
\affiliation{Panjab University, Chandigarh}
\affiliation{Peking University, Beijing}
\affiliation{Princeton University, Princeton, New Jersey 08545}
\affiliation{Saga University, Saga}
\affiliation{University of Science and Technology of China, Hefei}
\affiliation{Seoul National University, Seoul}
\affiliation{Sungkyunkwan University, Suwon}
\affiliation{University of Sydney, Sydney NSW}
\affiliation{Tata Institute of Fundamental Research, Bombay}
\affiliation{Toho University, Funabashi}
\affiliation{Tohoku Gakuin University, Tagajo}
\affiliation{Tohoku University, Sendai}
\affiliation{Department of Physics, University of Tokyo, Tokyo}
\affiliation{Tokyo Institute of Technology, Tokyo}
\affiliation{Tokyo Metropolitan University, Tokyo}
\affiliation{Tokyo University of Agriculture and Technology, Tokyo}
\affiliation{Toyama National College of Maritime Technology, Toyama}
\affiliation{University of Tsukuba, Tsukuba}
\affiliation{Virginia Polytechnic Institute and State University, Blacksburg, Virginia 24061}
\affiliation{Yonsei University, Seoul}
  \author{H.~Kakuno}\affiliation{Tokyo Institute of Technology, Tokyo} 
  \author{K.~Abe}\affiliation{High Energy Accelerator Research Organization (KEK), Tsukuba} 
  \author{K.~Abe}\affiliation{Tohoku Gakuin University, Tagajo} 
  \author{I.~Adachi}\affiliation{High Energy Accelerator Research Organization (KEK), Tsukuba} 
  \author{H.~Aihara}\affiliation{Department of Physics, University of Tokyo, Tokyo} 
  \author{Y.~Asano}\affiliation{University of Tsukuba, Tsukuba} 
  \author{T.~Aso}\affiliation{Toyama National College of Maritime Technology, Toyama} 
  \author{V.~Aulchenko}\affiliation{Budker Institute of Nuclear Physics, Novosibirsk} 
  \author{T.~Aushev}\affiliation{Institute for Theoretical and Experimental Physics, Moscow} 
  \author{A.~M.~Bakich}\affiliation{University of Sydney, Sydney NSW} 
  \author{Y.~Ban}\affiliation{Peking University, Beijing} 
  \author{S.~Banerjee}\affiliation{Tata Institute of Fundamental Research, Bombay} 
  \author{I.~Bizjak}\affiliation{J. Stefan Institute, Ljubljana} 
  \author{A.~Bondar}\affiliation{Budker Institute of Nuclear Physics, Novosibirsk} 
  \author{A.~Bozek}\affiliation{H. Niewodniczanski Institute of Nuclear Physics, Krakow} 
  \author{M.~Bra\v cko}\affiliation{University of Maribor, Maribor}\affiliation{J. Stefan Institute, Ljubljana} 
  \author{T.~E.~Browder}\affiliation{University of Hawaii, Honolulu, Hawaii 96822} 
  \author{Y.~Chao}\affiliation{Department of Physics, National Taiwan University, Taipei} 
  \author{B.~G.~Cheon}\affiliation{Sungkyunkwan University, Suwon} 
  \author{R.~Chistov}\affiliation{Institute for Theoretical and Experimental Physics, Moscow} 
  \author{S.-K.~Choi}\affiliation{Gyeongsang National University, Chinju} 
  \author{Y.~Choi}\affiliation{Sungkyunkwan University, Suwon} 
  \author{Y.~K.~Choi}\affiliation{Sungkyunkwan University, Suwon} 
  \author{A.~Chuvikov}\affiliation{Princeton University, Princeton, New Jersey 08545} 
  \author{S.~Cole}\affiliation{University of Sydney, Sydney NSW} 
  \author{M.~Danilov}\affiliation{Institute for Theoretical and Experimental Physics, Moscow} 
  \author{M.~Dash}\affiliation{Virginia Polytechnic Institute and State University, Blacksburg, Virginia 24061} 
  \author{L.~Y.~Dong}\affiliation{Institute of High Energy Physics, Chinese Academy of Sciences, Beijing} 
  \author{J.~Dragic}\affiliation{University of Melbourne, Victoria} 
  \author{A.~Drutskoy}\affiliation{Institute for Theoretical and Experimental Physics, Moscow} 
  \author{S.~Eidelman}\affiliation{Budker Institute of Nuclear Physics, Novosibirsk} 
  \author{V.~Eiges}\affiliation{Institute for Theoretical and Experimental Physics, Moscow} 
  \author{N.~Gabyshev}\affiliation{High Energy Accelerator Research Organization (KEK), Tsukuba} 
  \author{A.~Garmash}\affiliation{Princeton University, Princeton, New Jersey 08545} 
  \author{T.~Gershon}\affiliation{High Energy Accelerator Research Organization (KEK), Tsukuba} 
  \author{G.~Gokhroo}\affiliation{Tata Institute of Fundamental Research, Bombay} 
  \author{B.~Golob}\affiliation{University of Ljubljana, Ljubljana}\affiliation{J. Stefan Institute, Ljubljana} 
  \author{J.~Haba}\affiliation{High Energy Accelerator Research Organization (KEK), Tsukuba} 
  \author{C.~Hagner}\affiliation{Virginia Polytechnic Institute and State University, Blacksburg, Virginia 24061} 
  \author{T.~Hara}\affiliation{Osaka University, Osaka} 
  \author{M.~Hazumi}\affiliation{High Energy Accelerator Research Organization (KEK), Tsukuba} 
  \author{I.~Higuchi}\affiliation{Tohoku University, Sendai} 
  \author{L.~Hinz}{\affiliation{Swiss Federal Institute of Technology of Lausanne, EPFL, Lausanne}
  \author{T.~Hokuue}\affiliation{Nagoya University, Nagoya} 
  \author{Y.~Hoshi}\affiliation{Tohoku Gakuin University, Tagajo} 
  \author{W.-S.~Hou}\affiliation{Department of Physics, National Taiwan University, Taipei} 
  \author{H.-C.~Huang}\affiliation{Department of Physics, National Taiwan University, Taipei} 
  \author{T.~Iijima}\affiliation{Nagoya University, Nagoya} 
  \author{K.~Inami}\affiliation{Nagoya University, Nagoya} 
  \author{A.~Ishikawa}\affiliation{High Energy Accelerator Research Organization (KEK), Tsukuba} 
  \author{R.~Itoh}\affiliation{High Energy Accelerator Research Organization (KEK), Tsukuba} 
  \author{H.~Iwasaki}\affiliation{High Energy Accelerator Research Organization (KEK), Tsukuba} 
  \author{Y.~Iwasaki}\affiliation{High Energy Accelerator Research Organization (KEK), Tsukuba} 
% ...
\collaboration{The Belle Collaboration}

\date{\today}

\begin{abstract}
We report the measurement of an inclusive partial branching fraction for charmless semileptonic $B$ decay
and the extraction of $|V_{ub}|$.
Candidates for $B\rightarrow X_u \ell \nu$ are identified with a novel $X_u$-reconstruction method based on  neutrino reconstruction via missing 4-momentum and a technique called ``simulated annealing.''
Based on 86.9 ${\rm fb}^{-1}$ of data taken with the Belle detector, we obtain
${\Delta \cal B}(B \rightarrow X_u \ell \nu ; M_X<1.7\ \mathrm{GeV}/c^2, q^2>8.0\ \mathrm{GeV}^2/c^2)  = (7.37 \pm 0.89(\mathrm{stat.}) \pm 1.12(\mathrm{syst.}) \pm 0.55(b \rightarrow c) \pm 0.24(b\rightarrow u) ) \times 10^{-4}$
and determine 
$|V_{ub}| = (4.66 \pm 0.28(\mathrm{stat.}) \pm 0.35(\mathrm{syst.}) 
\pm 0.17(b\rightarrow c)\pm 0.08(b\rightarrow u)
 \pm 0.58(\mathrm{theory}))\times 10^{-3}$. 

\end{abstract}


\maketitle

The off-diagonal element $V_{ub}$ in the CKM matrix plays
an important role in $CP$-violation and rare decays of  the $B$ meson.
It is an important ingredient in overconstraining the unitarity triangle by measuring its sides and angles.
In the  experiments on the $\Upsilon(4S)$ resonance, its magnitude is  extracted from measurements of the 
$B \rightarrow X_u \ell \nu$ 
process in the limited  region of lepton momentum~\cite{CLEOendpt}  or the hadronic recoil mass $M_X$~\cite{BaBar}  where the contribution of background
from the $B \to X_c \ell \nu$ process is suppressed. These experiments achieve more precise measurements than LEP experiments~\cite{LEP} due to higher signal purity; however, the need to extrapolate measured rates from such limited regions results in large theoretical uncertainties on $|V_{ub}|$.
A recent theoretical development suggests that one can significantly reduce the theoretical uncertainty on the extrapolation by applying simultaneous cuts
on  $M_X$ and the invariant mass squared of
the lepton-neutrino system ($q^2$) in inclusive $B \rightarrow X_u \ell \nu $~\cite{bau}. 
We report here the first result with simultaneous requirements on $M_X$ and $q^2$.
The result is obtained with a  novel $X_u$-reconstruction method based on a combination of neutrino reconstruction and a technique called simulated annealing~\cite{anneal} to separate the two $B$ meson decays.
This method allows us to measure $M_X$ and $q^2$ with good efficiency so that it  achieves good statistical precision and small theoretical uncertainty with a modest integrated luminosity.
This analysis is based on $78.1\ {\rm fb}^{-1}$  data, corresponding to 85 million $B \bar{B}$ pairs, taken at the $\Upsilon(4S)$
resonance, and $8.8\ {\rm fb}^{-1}$ taken at an energy 60~MeV below the resonance, by the Belle detector
\cite{Belle} at the energy-asymmetric  $e^+ e^-$ collider KEKB~\cite{KEKB}.

We select hadronic events containing one lepton candidate (electron or
muon) having momentum above 1.2~GeV/$c$ in the center-of-mass-system (CMS) of the $\Upsilon(4S)$. 
To remove events with more than one neutrino, we exclude events containing
additional lepton candidates ($p^*_e > 0.5$~GeV/$c$ and $p^*_{\mu} > 0.8$~GeV/$c$). 
The neutrino is reconstructed
from the missing 4-momentum in the event
($\vec{p}_{\nu} \equiv \vec{p}_{\Upsilon(4S)} - \Sigma_i \vec{p}_i$,
$E_{\nu} \equiv E_{\Upsilon(4S)} - \Sigma_i E_i$). 
The net observed momentum ($\Sigma_i \vec{p}_i$) and energy ($\Sigma_i E_i$) are calculated using particles surviving track quality cuts based on the impact parameter to the interaction point and a shower energy cut.
Pairs of pions and electrons passing secondary vertex criteria
are treated as $K^0_S$ and photons, respectively.
All remaining charged tracks are classified as kaons, pions, or protons, based on  particle identification information.
$K_L^0$ candidates are identified from isolated clusters of hits in the detector for $K_L^0$s and muons~\cite{sin2phiprd}.
The energy of each particle candidate is calculated based on its momentum and mass assignment.
\begin{figure}[t]
\includegraphics[width=8.5cm,clip]{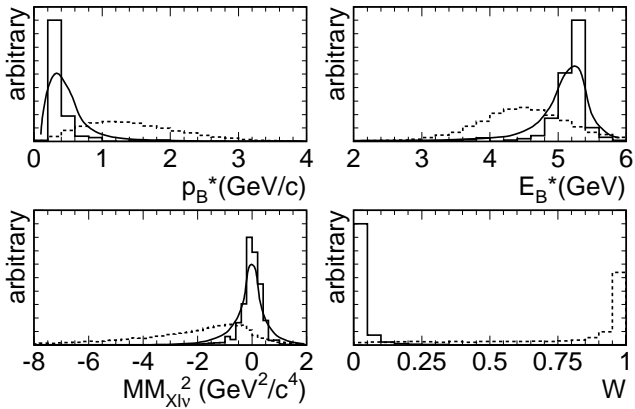}
\caption{Distributions for 3 discriminants and $W$ before (dashed histogram)  
and after (solid histogram) simulated annealing,  for real data. Distributions for 3 discriminants for correct combination of particles for MC events (solid curve). }
\label{discriminant}
\end{figure}
We then compute the missing mass of the event, defined as
$\mathit{MM}^2\equiv E_{\nu}^2/c^4-|\vec{p}_{\nu}|^2/c^2$, where the sign of $E_{\nu}^2$ is reversed when $E_{\nu} < 0$. 
We require  $-1.5\ {\rm GeV}^2/c^4<\mathit{MM}^2<1.5\ {\rm GeV}^2/c^4$
to suppress events with missing particles 
and with particles removed due to poor reconstruction quality.
For events that pass this requirement, 
we add back tracks and clusters rejected earlier due to reconstruction quality, selecting the combination that
gives the smallest value of $|\mathit{MM}^2|$. 
This determines the set of particles that are used in the subsequent analysis.
Events are further required to have a net
charge of 0 or $\pm 1$, and a polar angle for the missing momentum
within the barrel region ($32^{\circ} < \theta < 128^{\circ}$).
To suppress beam-gas events, we demand that the  net charge of all proton candidates be zero. 
Requiring that the
cosine of the angle of  $K^0_L$ candidates with respect to the
 missing momentum be less than 0.8 rejects events where the neutrino candidate
  is actually a $K^0_L$ meson. 

We then seek the most likely combination of particles belonging to $X \ell \nu$, the remainder
being from the associated $B$-meson ($B_{\rm opp}$).
Six discriminant variables are used: the momentum, energy, and polar angle of $B_{\rm opp}$ ($p_B^*$, $E_B^*$, $\cos \theta_B^*$) in the CMS, 
its charge
multiplicity ($N_{\mathrm{ch}}$) and net charge times the lepton charge ($Q_{B}\times Q_{\ell} $), and the missing-mass squared
recalculated with the energy and mass of $B_{\rm opp}$ constrained to the known values
($\mathit{MM}^2_{X \ell \nu}$).
Using Monte Carlo (MC) simulation events for $\Upsilon(4S) \to B \bar{B}$
where at least one $B$ decays into $X \ell \nu$, 
we determine  probability density functions (PDFs) 
for correct $X \ell \nu$ combinations
and for random $X \ell \nu$ combinations. 
Random candidates for $X \ell \nu$ consist of the lepton and neutrino candidates plus particles from the remainder of the event, selected randomly
so that the relative  multiplicities between $X$ and the remainder of the event  matches that at the generator level.
From the PDFs we calculate two likelihoods,
${\cal L}({\rm correct})$ and ${\cal L}({\rm random})$.
The most likely candidate combination in each event  is found by minimizing the parameter 
$ W \equiv {\cal L}({\rm random})/({\cal L}({\rm random}) + 
    {\cal L}({\rm correct}))$.

To minimize $W$, we have developed an approximate iterative algorithm based on simulated annealing.
We start from the initial candidate for $X \ell \nu$ that consists of the lepton and neutrino plus approximately one third of the remaining particles, selected randomly.
We move a random particle (other than the lepton or neutrino) between the $X$ and $B_{\rm opp}$ sides in an iterative way, where in one iteration we cross all particles at least once, and search for the combination 
that gives the minimum 
$W$ with 50 iterations. During the iteration process we take special care to reduce the chance of convergence to a local minimum of $W$.
\begin{figure}[t]
\includegraphics[width=8.5cm]{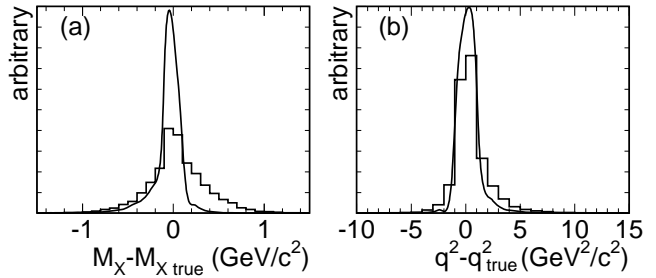}
\caption{$M_X$ (a) and $q^2$ (b)   resolution distributions for $B \to X_u \ell \nu$ MC events. 
Histograms (curves) show the results with the simulated annealing method
 (with correct particle assignment to $X_u$). }
\label{resol}
\end{figure}
\begin{figure}[htb]
\includegraphics[width=4.2cm]{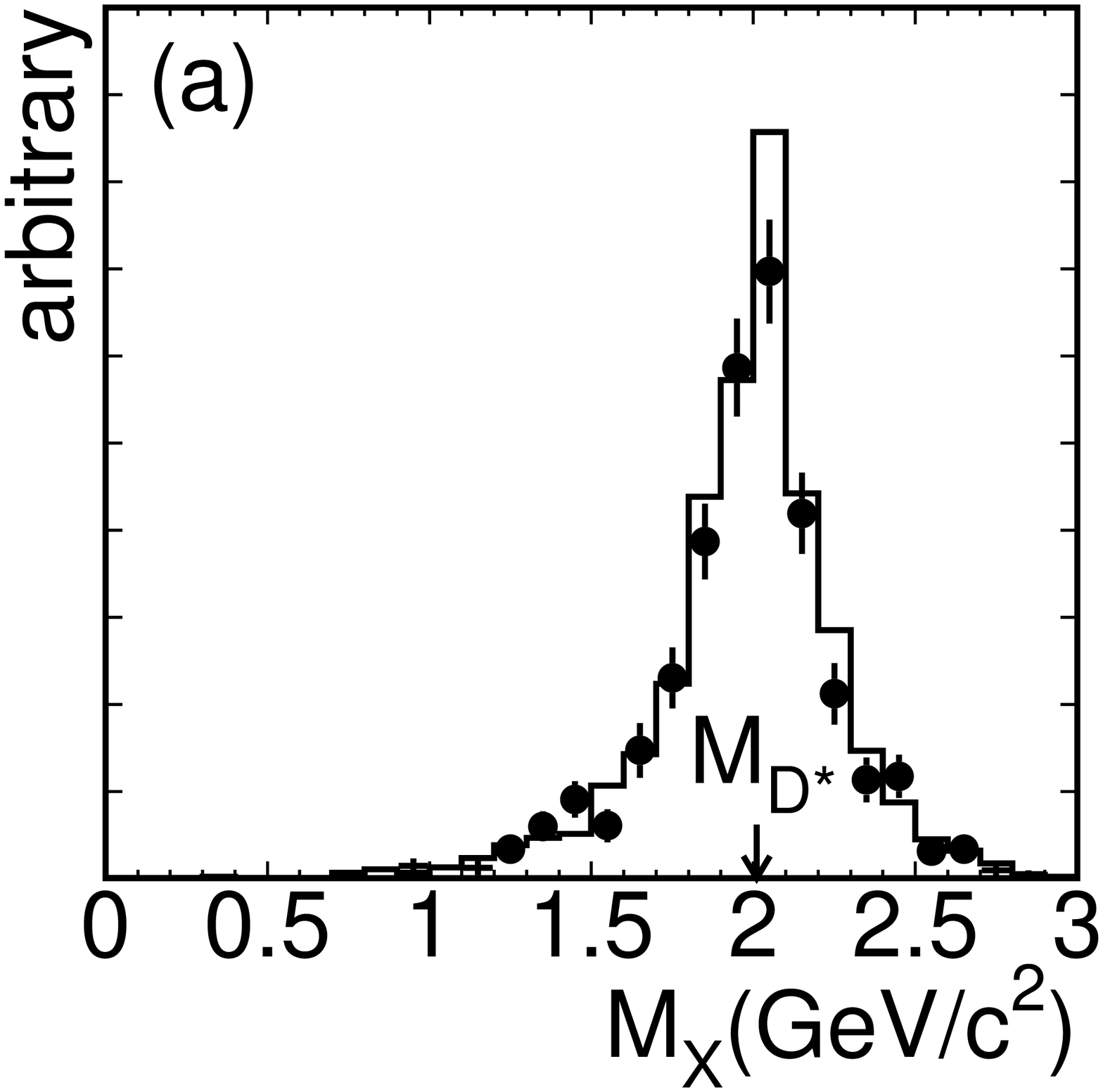}
\includegraphics[width=4.2cm]{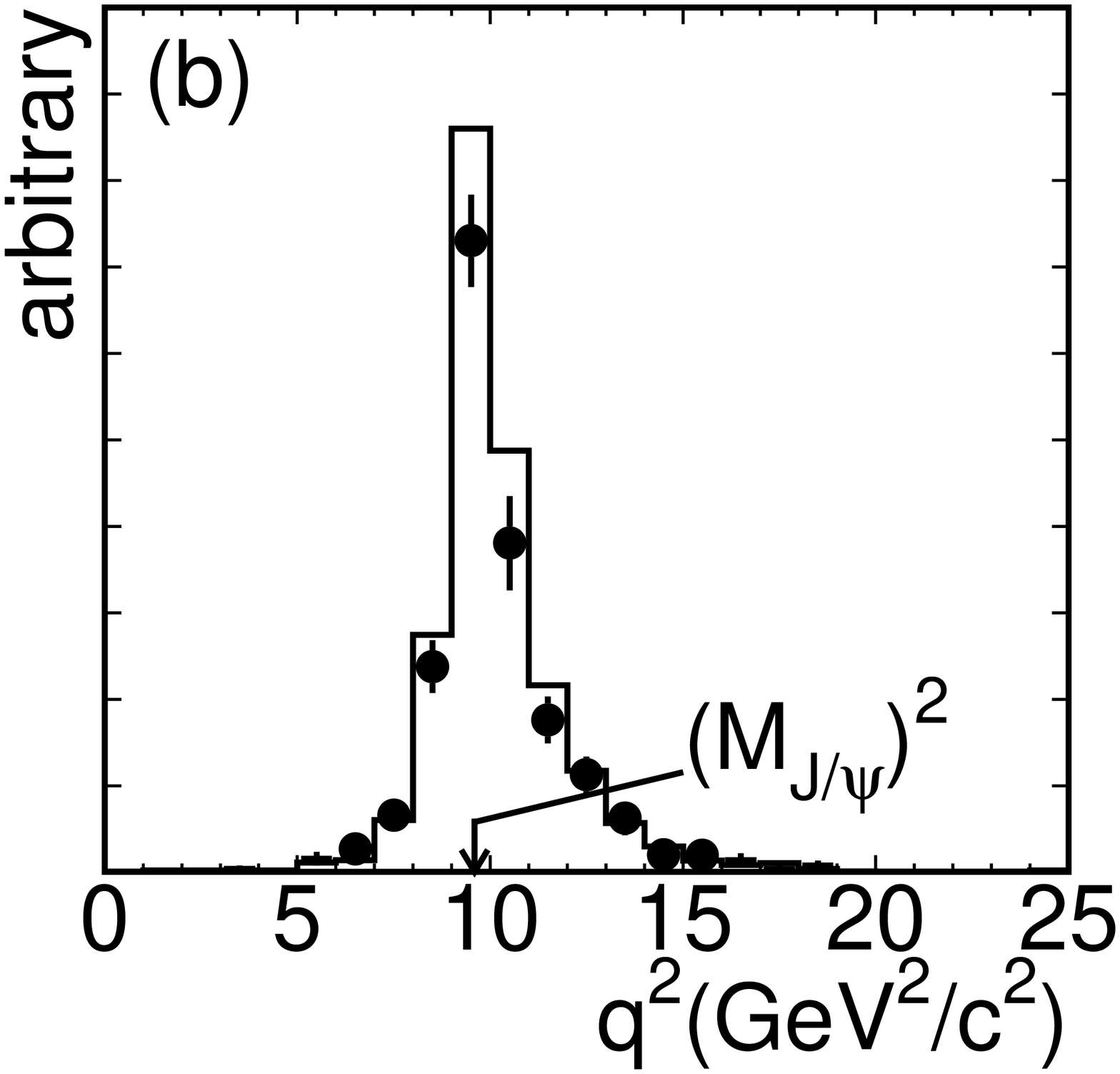}
\caption{(a) $M_X$ distributions for $B \to D^* \ell \nu$ control
 samples. 
 (b) $q^2$ distributions for $B \to J/\psi X$ control
 samples. Points are  data and the histogram is  MC.}
\label{control}
\end{figure}
 For instance, after every fifth iteration we compare
all combinations that can be constructed by crossing one particle and 
use the combination that gives the {\it largest} value of $W$ to seed a new cycle. 
We repeat this iteration process  10 times, starting each time with a different initial candidate, and select the case  with the smallest $W$.
Figure~\ref{discriminant} shows the
distributions in three of the six discriminant variables and $W$,  before and after simulated annealing. 
Also shown are the 
distributions for the correct combination in signal MC events.   

The final candidate is required to satisfy: 
i) $W < 0.1$, ii) $5.1 < E_B^* < 5.4$~GeV, iii) $0.25 < p_B^* <
0.42$~GeV/$c$, iv) $-2 < Q_{\ell} \times Q_{B} < +1$, and v) $-0.2 < \mathit{MM}^2_{X \ell \nu} < 0.4$~GeV$^2$/$c^4$. 
Contamination from the continuum is reduced  by demanding 
$|\cos \theta_{B\ell}|<0.8$, where $\theta_{B\ell}$ is the
angle between the thrust axis of $B_{\rm opp}$ and the lepton momentum. 
Figure~{\ref{resol}}   shows  the resolutions in $M_X$ and $q^2$  for $B \to X_u \ell \nu$ MC events.  Also shown in the figure are the resolutions for correct combination of particles.

The validity of the  method is checked with two data samples, one containing 38,600 fully-reconstructed $B \to D^* \ell \nu$ decays and the other containing
 84,100 $B \to J/\psi X, J/\psi\to\ell^+\ell^-$ decays.
These data samples are also used  to calibrate the detection efficiency.
For the $J/\psi X$ sample we treat one of the two leptons from $J/\psi$
as a neutrino, to emulate $B\to X \ell \nu$.
Corresponding MC events are generated with the QQ event generator~\cite{QQ}
 and the detector response is simulated using Geant~3~\cite{geant3}. 
Figure~{\ref{control}} shows the $M_X$ distribution for the $D^* \ell \nu$ sample and  the $q^2$ distribution for the $J/\psi X$ sample, with
MC distributions scaled to the number of background-subtracted events in the respective data samples. 
The peaks are at the $D^*$ mass and  $J/\psi$ mass squared, as expected, and the shapes are in good agreement between data and MC. 
Although these results verify that 
the simulated annealing method works as expected, 
we observe a small difference of efficiency between the data and MC.
By averaging the results of the two data samples, we obtain the efficiency 
ratio $r_{\rm{eff}}=0.891\pm0.043$ between the data and MC. 
\begin{figure} [t]
\begin{minipage}{4.25cm}
\includegraphics[width=4.65cm,clip]{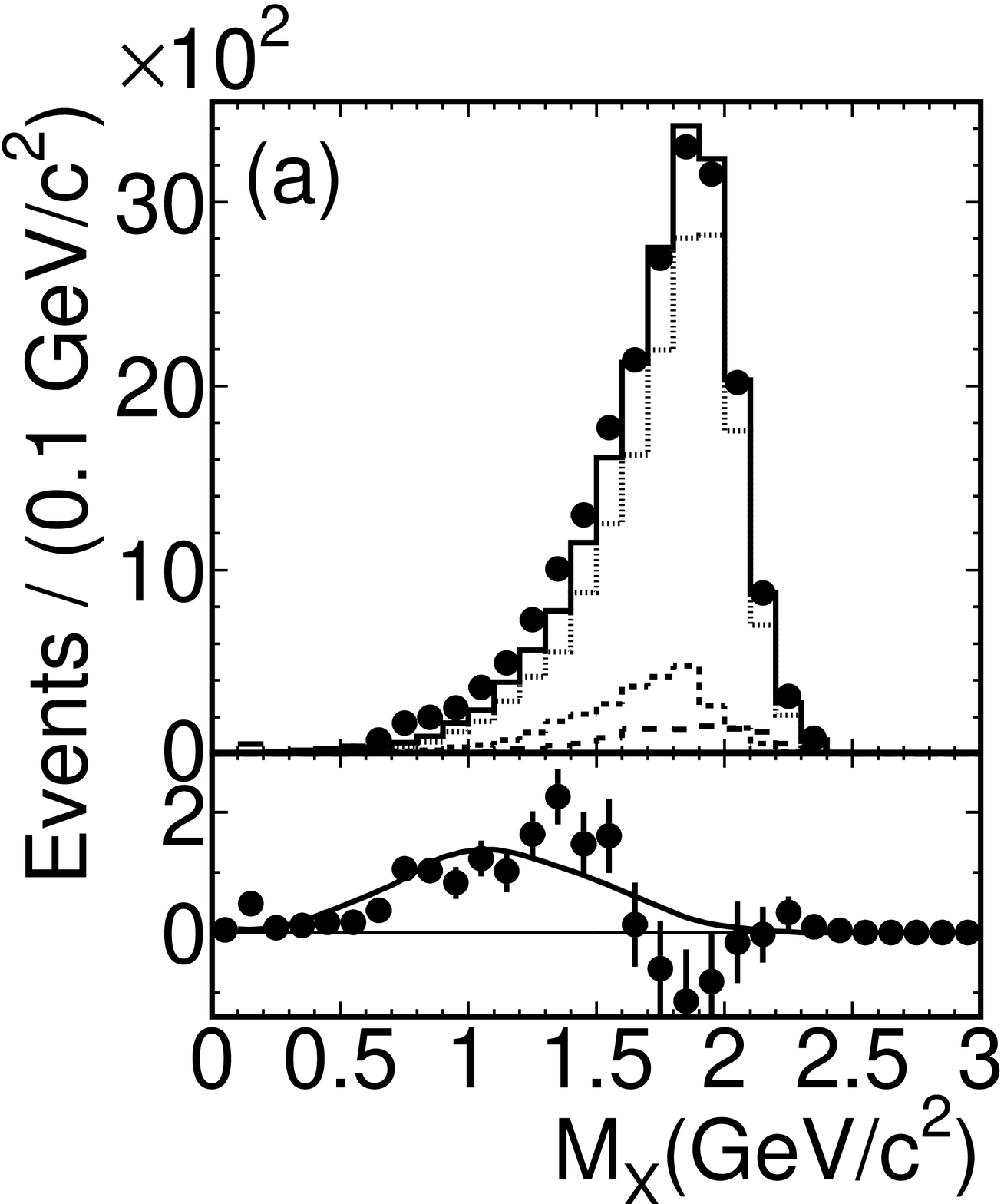}
\end{minipage}
\begin{minipage}{4.25cm}
\includegraphics[width=4.65cm,clip]{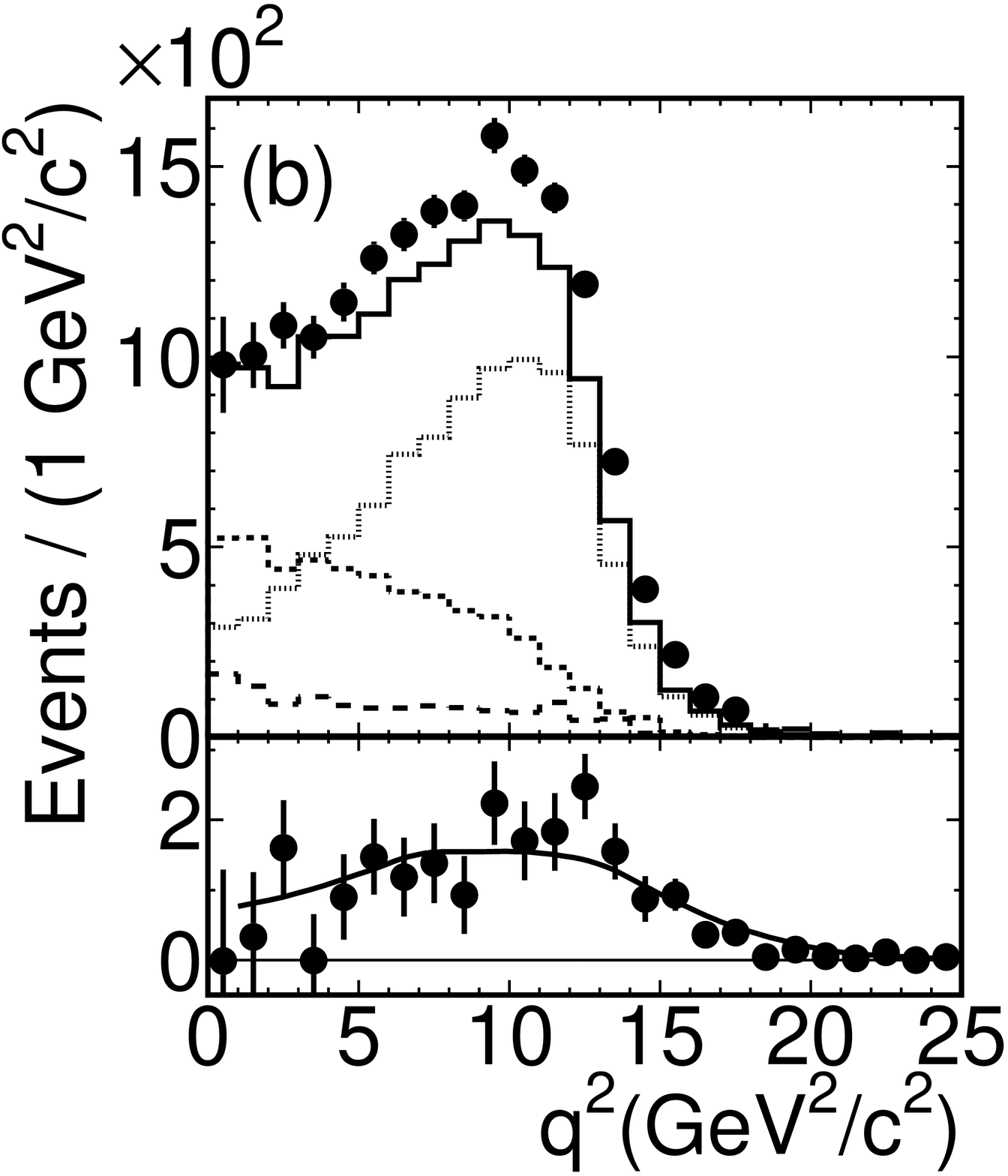}
\end{minipage}
\caption{(a) $M_X$ distribution for $q^2 > 8.0~\mathrm{GeV^2}/c^2$. 
(b) $q^2$ distribution for  $M_X<1.7\ \mathrm{GeV}/c^2$.
Points are the data and histograms are backgrounds from 
$D^* \ell \nu$ (dotted), $D \ell \nu$ (short dashed), others (long dashed), and total background contribution (solid). Lower plots show the data after background subtraction.
Solid curves show the inclusive MC predictions for $ B \to X_u \ell \nu$.}
\label{q2mx}
\end{figure}

We observe 8910 events in the $X_u \ell \nu$ ``signal'' region, defined as $M_X < 1.7 {\rm{GeV}}/c^2$ and $q^2 > 8.0 {\rm{GeV}}^2/c^2$.
These consist of semileptonic decays, $B\to X_{c,u} \ell
\nu$, other $B\bar B$ background, and residual continuum events.
The continuum contribution is estimated from off-resonance data to be $251\pm48$ events and is subtracted directly from the analyzed distributions.
The contributions from $B\to X_c \ell \nu$ and other $B\bar B$ backgrounds are estimated via MC in the ``background'' region $ M_X > 1.8\ {\mathrm{GeV}}/c^2$,  where  
$X_c \ell \nu$ dominates, and extrapolated to the signal region.
 We estimate them by fitting the $M_X$ and $q^2$ distributions from MC events  to those from the data using a two-dimensional $\chi^2$ fit method.
Contributions to  $X_c \ell \nu$ come from $D^{(*)} \ell \nu$, $D^{**} \ell \nu$,  and $D^{(*)} \pi \ell \nu$. 
Their branching fractions are floated in the fit.
The total rate for other $B\bar B$ backgrounds, 
arising from sources such as $b\rightarrow c \rightarrow s\ell\nu$ and fake leptons, which amounts to less than 1\% of events in the signal region,  
is floated.
The small $X_u \ell \nu$ contribution is estimated iteratively and is found to be $(0.94\pm 0.04)\%$ of events in the background region.
The obtained branching fractions for the exclusive $B\to X_c \ell \nu$ modes are consistent  with the PDG values~\cite{PDG2002}. 
The $B \bar{B}$ background in the signal region is estimated to be $7283\pm130\pm63$ events, where the first and second errors come from  fit and MC statistics, respectively.
The upper plots in Figure~\ref{q2mx}  show the $M_X$ distribution for $q^2 > 8.0\  {\rm{GeV^2}}/c^2$ and the $q^2$ distribution for  $M_X<1.7\ {\mathrm{GeV}}/c^2$ after continuum subtraction.
After subtracting the $B\bar B$ backgrounds, we obtain
 distributions for the $X_u \ell \nu$ signal,
shown in the lower plots. 
The net signal is $N_{\rm{obs}}$ = 1376$ \pm 167$ events where the error is statistical only.

In order to extract the partial branching fraction $\Delta \mathcal{B}$ for $B \to X_u \ell \nu$ in the signal region,
a Monte Carlo simulation is used to convert  $N_{\rm{obs}}$ to the  true number of signal events produced in this region, $N_{\rm{true}}$, and to estimate the efficiency for these events to be observed anywhere, $\epsilon_{\rm{signal}}$. In the MC simulation, $B \to X_u \ell \nu$ decays are simulated based on the prescription of ref.~\cite{fn}.
That analytic result gives  ${\cal O}(\alpha_s)$  corrections to leading order in the heavy-quark expansion for the triple differential $B \to X_u \ell \nu$ rate and includes the effect of the $b$-quark's Fermi motion. 
Two parameters therein, the $b$-quark pole mass, $m_b$, and the average momentum squared of the $b$-quark inside the $B$ meson, $\mu_{\pi}^2$, are derived from the CLEO measurements of the hadronic mass moments in inclusive $B \to X_c \ell \nu$ and photon energy spectrum in $B \to X_s\gamma$~\cite{lambda}.
We use $m_b = 4.80\pm0.12\, {\rm{GeV}}/c^2$ and $\mu_{\pi}^2 =0.30\pm0.11\,{\rm{GeV}}^2/c^2$, which differs from CLEO's evaluation in that terms proportional to $1/m_b^3$ and $\alpha_s^2$ have been removed from the relation between the measured observables and $m_b$ and $\mu_{\pi}^2$.
 The MC events are generated with the EvtGen generator~\cite{evtgen}.
$ N_{\rm{true}}$ is estimated by
$ N_{\rm{true}} = N_{\rm{obs}} \times F (F = 1 + N_2/N_1 -N_3/N_1)$.
Here $N_1$ is the number of events observed in the signal region and $N_2$ ($N_3$) is  the number of events generated inside (outside) the signal region and observed outside (inside) the signal region.
We find $F=0.938$, and thus
$N_{\rm{true}} = 1291\pm 157$. 
The efficiency   $\epsilon_{\rm{signal}}$ is predicted to be 0.578\%.
We  determine $\Delta \mathcal{B}$  by
$0.5\times N_{\rm{true}}/(\epsilon_{\rm{signal}}\times r_{\rm{eff}})/(2N_B)$,
where $r_{\rm{eff}}$ is the efficiency correction factor described earlier, $N_B$ is the number of $B\bar{B}$ events
and  the factor  0.5 is needed to take into account the electron and muon
data:
\begin{eqnarray}  
 \Delta\mathcal{B} = (7.37\pm0.89\pm1.12 \pm 0.55 \pm0.24)\times10^{-4}. \nonumber
\end{eqnarray}
The errors are statistical, systematic, from 
$B \to X_c \ell \nu$ model dependence, and $B \to X_u \ell \nu$ model dependence, respectively.
Sources of  systematic uncertainty include 
signal MC statistics (1.8\%), lepton identification (2.6\%),   
uncertainty of $F$ due to imperfect detector simulation (1.2\%),
selection and reconstruction efficiency (4.9\%),
$B\bar{B}$ background estimation (14.0{\%}).
The uncertainty in the $B \bar{B}$ background estimation is from 
MC statistics (4.6{\%})
and distortion of the $M_X$ and $q^2$ distributions due to imperfect detector 
simulation (13.2{\%}).  
Major contributions to the last error are from $K^0_L$ contamination (8.6{\%}),
 electromagnetic cluster finding efficiency (8.2{\%}), 
lepton efficiency (3.2{\%}), 
clusters produced by charged tracks (2.9{\%}), 
lepton fake rate (2.7{\%}), 
and $K/\pi$ separation (2.7{\%}).
The error from $K^0_L$ contamination is estimated using inclusive $K^0_S$ events 
where we discard  $K^0_S$s to emulate inclusive $K^0_L$ events.
The error from the cluster finding efficiency is estimated by reducing the photon-finding efficiency within its uncertainty.
The  model dependence of $X_c \ell \nu$ is estimated to be 7.4{\%} by 
 varying  
the $D_1 \ell \nu$ plus $D^*_2 \ell \nu$ fraction in the $D^{**}\ell \nu$
by 25{\%} and by varying  the slope parameters of the form factors for $D \ell \nu$ and $D^* \ell \nu$,  $\rho_D^2=1.19\pm0.19$ and $\rho^2=1.51\pm0.13$ ~\cite{PDG2002}, within their errors.
$B \to X_u \ell \nu$  model dependence  (3.4{\%}) is estimated by varying the parameters of the inclusive model within their errors and by comparing to a
simulation with a full exclusive implementation of the ISGW2 model~\cite{ISGW2}.

In the context of HQET and OPE the partial branching fraction $\Delta\mathcal{B}(B \to X_u \ell \nu)$ is related to $|V_{ub}|$~\cite{bau,HLM,ligeti},
\begin{eqnarray*}  
|V_{ub}| = 0.00444 \left(  \frac  {\Delta \mathcal{B}(B \rightarrow X_u \ell \nu )}
                       {0.002\times 1.21G(q^2_{\rm{cut}},m_{\rm{cut}})} 
                             \frac{1.55\mathrm{ps}}
                                  {\tau_B}   
                              \right)^{1/2}
\end{eqnarray*}
where 
$1.21G(q^2_{\rm{cut}},m_{\rm{cut}})= f_{\rm{HQET}}\times \left( \frac {m_b^{\rm{1S}}}{4.7{\rm{GeV}}/c^2}\right)^5 $,  $f_{\rm{HQET}}$ represents the fraction of events with $q^2>q^2_{\rm{cut}}$ and $M_X<m_{\rm{cut}}$, and 
$m_b^{\rm{1S}}$ is one-half of the perturbative contribution to the mass of the $\Upsilon(1S)$.
$G(q^2_{\rm{cut}},m_{\rm{cut}})$ is calculated 
to ${\cal O}(\alpha_s^2)$ and ${\cal O}(1/m_b^2)$ in {\cite{bau}}, including the effect of the Fermi motion of the $b$ quark,
 which is expressed in terms of $m_b^{\rm{1S}}$.
We use $m_b^{\rm{1S}}=4.70\pm0.12\,{\rm{GeV}}/c^2$~\cite{bau,mb1s}, which gives
$G(q^2_{\rm cut},m_{\rm cut})=0.268$ \cite{bau,ligeti}.
 The theoretical uncertainty on $|V_{ub}|$ is determined only by the uncertainty on $G(q^2_{\rm{cut}},m_{\rm{cut}})$.
The uncertainty on  $G(q^2_{\rm{cut}},m_{\rm{cut}})$,  in total 25$\%$,  consists of 6{\%} for perturbative, 8{\%} for nonperturbative terms (dominated by the weak annihilation contribution),  and 23{\%} from  the uncertainty on $m_b^{\rm{1S}}$~\cite{bau,duality}.
The 23\% error contains 10\% for 
$ f_{\rm{HQET}}$ and 13\% for  $(m_b^{\rm{1S}})^5$. 
These uncertainties are positively correlated, so we add them
linearly, whereas they have been given separately in conventional
analyses.
Using $\tau_B =1.604\pm0.012$ ps~\cite{PDG2002},
we obtain 
\begin{eqnarray}  
|V_{ub}| &= & (4.66 \pm 0.28 \pm0.35 \pm 0.17 \pm 0.08 \pm  0.58) \times
10^{-3} \nonumber
\end{eqnarray}  
where the errors are statistical, systematic,  $b \to c$ model dependence, 
$b \to u$ model dependence, and theoretical uncertainty for OPE, respectively.

To summarize, we have performed the first measurement of $|V_{ub}|$ with simultaneous requirements on $M_X$ and $q^2$ using a novel $X_u$-reconstruction method. The result of $|V_{ub}| = (4.66 \pm 0.76) \times 10^{-3}$ is consistent with the previous inclusive measurements~\cite{LEP,CLEOendpt,BaBar}  and the  total error  is comparable with those of the  previous measurements on $\Upsilon(4S)$~\cite{CLEOendpt,BaBar}.
Due to simultaneous requirements on $M_X$ and $q^2$, the $f_{\rm{HQET}}$ error is much smaller than those  of the previous measurements on $\Upsilon(4S)$~\cite{CLEOendpt,BaBar}.

We wish to thank the KEKB accelerator group for the excellent
operation of the KEKB accelerator.
We acknowledge support from the Ministry of Education,
Culture, Sports, Science, and Technology of Japan
and the Japan Society for the Promotion of Science;
the Australian Research Council
and the Australian Department of Education, Science and Training;
the National Science Foundation of China under contract No.~10175071;
the Department of Science and Technology of India;
the BK21 program of the Ministry of Education of Korea
and the CHEP SRC program of the Korea Science and Engineering Foundation;
the Polish State Committee for Scientific Research
under contract No.~2P03B 01324;
the Ministry of Science and Technology of the Russian Federation;
the Ministry of Education, Science and Sport of the Republic of Slovenia;
the National Science Council and the Ministry of Education of Taiwan;
and the U.S.\ Department of Energy.

\end{document}